\documentclass[10pt]{iopart}
\usepackage{iopams,amsfonts}
 \usepackage{epsfig}

\def\a{\alpha}
\def\b{\beta}

\def\d{\delta}

\def\vt{\vartheta}

\def\A{{\cal{A}}}

\def\a{\alpha}
\def\b{\beta}

\def\d{\delta}

\def\hi{{\hat{i}}}
\def\hj{{\hat{j}}}
 \def\hk{{\hat{k}}}
 \def\hm{{\hat{m}}}
 \def\hn{{\hat{n}}}
\def\L{{\mathcal{L}}}
\def\T{{\mathcal{T}}}

\def\F{{\mathcal{F}}}
 \def\M{{\mathcal{M}}}
 
\def\G{\Gamma}
\def\oG{\stackrel{{\rm o}}{\Gamma}{\!}}
 \def\oT{\stackrel{{\rm o}}{T}{\!}}
\def\LG{\stackrel{{\rm *}}{\Gamma}{\!}}
 \def\brr{\begin{eqnarray}}
\def\err{\end{eqnarray}}
\def\brn{\begin{eqnarray*}}
\def\ern{\end{eqnarray*}}
\eqnobysec

\begin{document}

\title{Maxwell-type  behavior from a  geometrical structure}

\author{Yakov Itin}

 \address{Institute of Mathematics,
Hebrew University of Jerusalem 
 Givat Ram, Jerusalem 91904, Israel\\
 Jerusalem College of Technology, Jerusalem 91160 Israel}
 \ead{itin@math.huji.ac.il}


\begin{abstract}
 We study  which geometric structure can be constructed 
 from the  vierbein  (frame/coframe) variables  and which field models 
 can be related to this geometry.
 The coframe field models, alternative to GR, are known as
viable models for gravity, since they have  the
Schwarzschild solution. Since the local Lorentz invariance is 
 violated, a physical interpretation of
additional six degrees of freedom is required. The
geometry of such models is usually given by two different
connections --- the Levi-Civita symmetric and
metric-compatible connection and the Weitzenb\"{o}ck flat
connection.

 We  construct a general family of linear connections
 of the same type,
 which includes two connections above  as
 special  limiting cases.
 We show that for dynamical propagation of six additional
 degrees of freedom it is  necessary for the
 gauge field of infinitesimal transformations
 (antisymmetric tensor)
 to satisfy the system of two first order differential
 equations. This system is similar to the vacuum Maxwell
 system and even coincides with it on a flat manifold.
 The corresponding ``Maxwell-compatible connections'' are
  derived.
 Alternatively, we derive the same Maxwell-type  system
 as a symmetry conditions of the viable models Lagrangian.
 Consequently we derive a nontrivial decomposition of
 the coframe field to the pure metric field plus a
 dynamical field of infinitesimal Lorentz rotations.
 Exact spherical symmetric solution for our dynamical field
  is derived. It is bounded near
 the Schwarzschild radius. Further off, the solution is close to the
 Coulomb field.

\end{abstract}
\date{\today}
\section{Introduction}
GR is a  classical field theory for 10 independent
variables --- the components of metric tensor $g_{ij}$. It is well
known, however, that some problems inside and beyond Einstein's gravity
 require a richer set of 16 independent variables
--- the components of the coframe (aka reper, vierbein, ...).
In the following issues of gravity, the coframe is not
only a useful tool but often cannot even be replaced
by the standard metric variable:
\begin{itemize}
\item[(i)] Hamiltonian formulation \cite{Ashtekar:1987gu},\cite{Deser:1976ay} ;
\item[(ii)] Positive energy proofs \cite{Nester:1994du};
\item[(iii)] Fermions on a curved manifold \cite{Deser:1974cy},\cite{spin};
\item[(iv)] Supergravity \cite{VanNieuwenhuizen:1981ae};
\item[(v)] Loop quantum gravity \cite{Kummer:2005tx}, \cite{Perez:15jz}.
\end{itemize}

 Absolute (teleparallel) frame/coframe variables  were
 introduced in physics by Einstein in 1928 with an aim of a unification 
 of gravitational and electromagnetic fields (for classical references, 
 see \cite{cit1}). 
In GR description, the additional six degrees
of freedom do not have a physical sense and treated as  a type 
 of a  gauge symmetry of the metric tensor.
 
It was already noticed by Einstein,  that 16 reper
components cannot be completely equivalent to 10 components 
 of the metric tensor. 
 Indeed, the supergravity and the loop quantum gravity models apparently 
 can be formulated only in term of vierbein. 
 So, it is natural to study which geometric structure can be constructed 
 from the  vierbein  (frame/coframe) variables  and which field models 
 can be related to this geometry. This is a subject of the current paper.

 The frame  field $e_\a$ and its dual, the coframe field $\vt^\a$,
 have a well defined geometrical sense. In particular, even been
 considered as independent physical fields,
 they provide a special  (absolute) reference basis.
 For the  bases, $\{e_\a\,,\vt^\a\}$, fixed at a point,
 the construction gives an invariant
 meaning to the components of a tensor, thus it emerges in violation
 of the rotational and Lorentz invariance \cite{Bluhm:2004ep}.
 However, when the {\it global} (rigid) 
 Lorentz transformations of the absolute basis fields
 are acceptable, the frame components of a tensor are
 transformed merely by the Lorentz transformation law.
 Thus, some interrelation between the Lorentz
 invariant field theories
 and the diffeomorphism invariant gravity emerges.
 When the absolute basis field is restricted to  a point, the
 Lorentz invariance requires to consider it   not alone but  as
 a member of a class of equivalent bases. The equivalence
relation for this class is provided by a group $G$ of transformations
$\vt^\a\to L^\a{}_\b\,\vt^\b$, which has to include local physics symmetries,
 i.e., the rotations and the boosts.

 When we are dealing with the absolute basis fields on a  manifold,
 the relation between the frames at
 different points is not governed by the constrains of local
 physics. Consequently, three principally different possibilities are open:

(i) {\it Riemannian geometry.} The frames at distinct  points are
not related to one  another at all, i.e. the   dependence of the elements
 of the group $G$ on a point is
absolutely arbitrary. It is possible only if the corresponding geometry,
i.e the metric and the connection,  respect the arbitrary
transformations of the frame field. The unique connections with
this property is the {\it Levi-Civita connection} of Riemannian
geometry. This case is realized in the standard GR when it is
formulated in an arbitrary non-holonomic frame. The frame/coframe
fields are not physical in this case and only play a role of a
useful reference tool. They  can be replaced even by holonomic
bases generated by local coordinates.

(ii) {\it Teleparallel geometry.} Another limiting case emerges
when the frames in distinct points are  strongly connected to one
 another. It means that when a frame in an arbitrary point is rotated
in a certain angle, the
frames at distant points immediately rotated in the same angle.
The unique connection with this property is the
flat  {\it Weitzenb\"{o}ck connection.} Although the scalar curvature is
zero for such a connection, the action functional can be
constructed from the torsion tensor. It is natural to require
this action functional to
be invariant only under global transformations of the basis
field. The analysis of this model shows that it has not spherical
symmetric solutions with Newtonian behavior at infinity. As a
result, this absolute teleparallel gravity model is not physical.

(iii) {\it Gauge geometry.}
 Besides the two limiting cases given above, there is a family of  geometries
 with the field of frame rotations satisfying some system of
 differential equations. The basic 
 geometrical quantities, metric and connection, have to be invariant under
 local transformations of the frame field satisfied these equations.
 In this construction, the frame field does not
 emerge explicitly, so we will refer to the corresponding
 geometrical structure as a {\it gauge coframe geometry}.

In alternative gravity models (see \cite{Hayash:1979} --- 
 \cite{Obukhov:2002tm} and the reference given therein), 
 the frame variable appears as an independent  physical field.
 In the first order approximation \cite{Itin:2004ig},  the coframe field
 is separated to a sum of two independent fields - a metric field and an
 antisymmetric field of the rang two. The same separation emerges in
 the Lagrangian, in the field equation and also in the energy-momentum
 tensor.  In the current paper we show that  a geometrical and
 physical interpretation of the antisymmetric part can be prolongated
 in the higher order approximations. 
 Our main result is: In the viable coframe models,
 alternative to GR, the additional degrees of freedom
 can be interpreted as a new dynamical field which behavior
 in the first order approximation is described by
 the vacuum Maxwell-type system.

\subsection{Overview}

  The aim of this paper is to derive
 the mentioned coframe gauge geometrical structure and
 its possible applications to classical fields.
 The paper is organized as followed.

 In section 2, we construct a
 most general class of connections which are linear
 in the first
 order derivatives of the coframe field.
 This six-parametric family
 involves the Levi-Civita and the  Weitzenb\"{o}ck
 connections as special limiting cases.
 We also identify the sub-families
 of torsion-free and metric compatible connections.

  In section 3, the behavior of the coframe
  connections
 under local $SO(1,3)$ transformations of the coframe is considered.
 Besides the
 Levi-Civita and the  Weitzenb\"{o}ck connections,
 we identify a sub-family of gauge invariant connections.
 The  corresponding constraints  compose a system of 8
 first order partial differential equation for six
 independent entries of a $SO(1,3)$ matrix. This situation
 is very similar to the standard Maxwell system of eight
 field equations for six independent components of the
 electromagnetic field.

 In section 4, we derive the same system of constrains
 from a different (physical) point of view.
 We require a most general quadratic
 coframe Lagrangian to be invariant under $SO(1,3)$
 transformations of the
 coframe. The gauge invariant Lagrangians turn out to
 be in a correspondence
 with the known viable coframe models having the
 Schwarzschild solution.

 In section 5, we study  the first order approximation
 to the system of the constrains.
 On a flat manifold, this system coincides with the
 vacuum Maxwell equations.
 On a curved manifold, it turns out to be a system of
 covariant Maxwell-type equations when the covariant
 derivatives are considered relative
 to the  Weitzenb\"{o}ck connections.

 In section 6, we derive an exact spherical symmetric
 solution to our system of constrains. We also compare
 our model with the standard description of interaction
  between gravitational
and electromagnetic fields. In contrast to the standard
Einstein-Maxwell
 system,  our model predicts  mass dependence in the
 Coulomb-type law. We derive
 the exact expression of the corresponding correction
 term.

 In section 7, we give an outlook of the proposed alternative model
 and discuss how (and if) the additional degrees of freedom can be 
 related to the
 ordinary electromagnetic field.
\subsection{Notations}
We use the Greek indices  $\a\,, \b\,, \cdots=0,1,2,3$ to identify the
specific vector fields of the frame $e_\a$ and the specific 1-forms of the
coframe $\vt^\a$.  The Roman indices $i\,,j\,, \cdots=0,1,2,3$ refer to
 local coordinates. Summation from 0 to 3 is understood
over repeated indices of both types (Einstein's summation
convention). Two types of indices   and basically different.
 In particular, they cannot be summed (contracted)
in  the expressions ${\vt^\a}_a$ or ${e_\a}^{a}$.
The Lorentz metric is used with the  sign agreement
$\eta_{\a\b}={\rm diag} (-1,1,1,1)$.
 The spatial
indices are denoted as $\hi\,,\hj\,,\cdots=1,2,3$.

 We denote the coefficients of connection as $\G_{ij}{}^k$ and $\G_{ijk}=
 g_{km}\G_{ij}{}^m$.
 Such notation is useful for the exterior form representation \cite
 {Hehl:1994ue}. In order to go back to the ordinary tensorial
 notations  $\G^k_{ij}$, it is enough to move the last index to the first
 position.

 The symmetrization and antisymmetrization operators are used in
 the normalized form, i.e., $A_{(i_1\cdots i_p)}=( 1/{p!})(A_{i_1\cdots i_p}+
 \cdots)$ and $A_{[i_1\cdots i_p]}=( 1/{p!})(A_{i_1\cdots i_p}-
 \cdots)$.

 \section{Coframe connections}
\subsection{Coframe structure}
Let a  $4D$-manifold be endowed with a  frame field
$e_\a$ and a  coframe field $\vt^\a$.
 All the structures are assumed to be smooth almost everywhere.
 In local coordinates
$x^i$, the  fields  are expressed respectively as
\begin{equation}\label{frame-def}
 e_\a={e_\a}^{i}\,\partial_i \,,\qquad \vt^\a={\vt^\a}_i \,dx^i\,,
\end{equation}
i.e., by  two $4\times 4$ matrices ${e_\a}^{i}(x)$ and ${\vt^\a}_i(x) $ of
 smooth functions.
 These two basis fields are  assumed to be reciprocal to each other:
\begin{equation}\label{recip}
{e_\a}^{i}\,{\vt^\b}_i=\d^\b_\a\,, \qquad
{e_\a}^{i}\,{\vt^\a}_j=\d^i_j\,.
\end{equation}
So, in fact,  only one field, $e_\a$ or $\vt^\a$, is an
 independent variable.
We prefer to use the coframe field $\vt^\a$ as a basis variable
since  it is suitable for  a compact  exterior form
representation.

For a rigidly fixed coframe field, the components of an arbitrary tensor
 obtain an invariant sense when referred to the special bases $e_\a\,,\vt^\a$.
 It emerges in hard violation of Lorentz invariance.
 Thus, in order to have a Lorentz invariant field model, the
 coframe field has to be defined only up to  global
 Lorentz transformations.
The gauge paradigm requires to localize the global transformations,
 so we define a {\it coframe structure} as a triplet
\begin{equation}\label{str1}
 \{\M\,,  \, \vt^\a\,,  \, G\}\,,
\end{equation}
 where $\M$ is a smooth manifold,
 $\vt^\a$ is a smooth coframe
 field on $\M$, and $G=\{L^\a{}_\b(x)\}$ is a local (pointwise) group
 of transformations
 $\vt^\a\to L^\a{}_\b(x) \vt^\b$.
 Although the coframe variable  itself is a pure geometrical object,
it is not clear  what geometry is generated by the structure (\ref{str1})
 for different groups $G$.

 We accept the Cartan viewpoint that treats a
 geometrical structure as a pair of two
 independent objects:
 a metric tensor $g_{ij}(x)$ and an asymmetric connection
 field $\G_{ij}{}^k(x)$.
 For the specific coframe structure (\ref{str1}),
 we require both fields
 to be explicitly constructed from the coframe components.
 In other words, we specify (\ref{str1}) to
 \begin{equation}\label{str2}
 \{\M\,, \, g_{ij}(\vt^\a)\,,  \,  \G_{ij}{}^k(\vt^\a)\,, \,  G\}\,.
\end{equation}
In correspondence to  local physics, the metric tensor has to
be  defined as
  \begin{equation}\label{metr}
g=\eta_{\a\b}\,\vt^\a\otimes \vt^\b\,,
 \end{equation}
 or, in components,
 \begin{equation}\label{metr1}
 g_{ij}=\eta_{\a\b}\,\vt^\a{}_i\,\vt^\b{}_j\,,\qquad
 g^{ij}=\eta^{\a\b}\,e_\a{}^i\,e_\b{}^j\,.
\end{equation}
Here $\eta_{\a\b}$ is the Lorentzian metric in the tangential
vector space. Indeed, in a small neighborhood of a point, the coframe
 field is approximately  holonomic $\vt^\a{}_i=\d^\a_i$,
 so  (\ref{metr1}) is locally approximated by Lorentzian metric,
 see \cite{Itin:2004ig}.
 Due to the index content, (\ref{metr})  is  a unique construction
 up to a constant scalar factor.  This factor can be neglected by rescaling
 the coframe field (recall that the global $GL(4,\mathbb R)$-transformations
 of the structure are  admissible).
 Hence (\ref{metr}) is a unique construction of
 the metric tensor from the coframe components when the
 conformation with the
 local Lorentz metric is required. This definition restricts
 the freedom of the coframe transformation to local pseudo-rotations,
 i.e., at every point $G=SO(1,3)$.
 The entries of this $SO(1,3)$-matrix remain arbitrary functions of a point.

\subsection{Coframe connections}
Let a manifold be endowed with a field of asymmetric Cartan connections.
Relative to a local coordinate chart
 $x^i$, every connection is represented by a set of $4^3$ independent
 functions ${\G_{ij}}^k(x)$ ---  the coefficients of the
 connection.
 The only condition the functions ${\G_{ij}}^k(x)$ have to satisfy is to
   transform, under a change of coordinates $x^i\mapsto y^i(x^j)$,
   by an inhomogeneous linear rule:
\begin{equation}\label{cc1}
\G_{ij}{}^k\mapsto \Big(\G_{lm}{}^n y^l{}_{,i} y^m{}_{,j}+
y^n{}_{,ij} \Big)  x^k{}_{,n}\,,
 \end{equation}
 where the derivatives are denoted as
 $y^i{}_{,j}={\partial y^i}/{\partial x^j}$ and
 $x^i{}_{,j}={\partial x^i}/{\partial y^j}$.

For the coframe connections,  we require $\G_{ij}{}^k$ and
 $\G_{ijk}=g_{km}\G_{ij}{}^m$ to be explicitly constructed from the  coframe
 components and their derivatives. Moreover, we require $\G_{ij}{}^k$
 to be linear in  $\vt^\a{}_{m,n}$ --- linear connection.
 This requirement is in parallel to the  standard construction of
 Riemannian geometry. Indeed,   the Levi-Civita connection can be treated
 as a unique linear connection which can be constructed from
 the first order derivatives of the metric components $g_{ij,k}$.

 Two examples of the coframe connection satisfied the requirement  above
  are well known:

 (i) The  Weitzenb\"{o}ck connection is defined as
  \begin{equation}\label{weiz2}
 \oG_{ij}{}^k={e_\a}^k\,{\vt^\a}_{i,j}\,, \qquad
 \oG_{ijk}=\eta_{\a\b}\vt^\a{}_k\vt^\b{}_{i,j}\,.
 \end{equation}
 It is straightforward to check the transformation rule
 (\ref{cc1}) for this expression. The Riemann curvature of this
 connection is identically zero, so the connection is flat.
 We  denote the  antisymmetric  combination (torsion)
 of (\ref{weiz2}) and its trace as
 \begin{equation}\label{C-def}
 C_{ijk}=\oG_{[ij]k}\,,\qquad C_i=C_{mi}{}^m\,.
 \end{equation}
 When the transformation (\ref{cc1}) is applied to these quantities,
 the inhomogeneous parts are canceled. Consequently,
 (\ref{C-def}) change as tensors under the transformations of coordinates.
 Their behavior under local transformations of the coframe is an independent
 property, which we will examine in the consequence.

 (ii) The Levi-Civita connection of  Riemannian
geometry,
\begin{equation}\label{LC-con}
 \LG_{ij}{}^k=\frac 12 g^{km}(g_{im,j}+g_{jm,i}-g_{ij,m})\,,
 \end{equation}
 can be rewritten as a linear combination of the first order
 derivatives of the coframe components.
 Indeed, we substitute (\ref{metr}) into (\ref{LC-con})
 to have
\begin{eqnarray}\label{LC1}
 \LG_{ijk}&=&\eta_{\a\b}\Big(\vt^\a{}_k\vt^\b{}_{(i,j)}+
 \vt^\a{}_j\vt^\b{}_{[k,i]}+
 \vt^\a{}_i\vt^\b{}_{[k,j]}\Big)\nonumber\\
&=& \oG_{(ij)k}+\oG_{[ki]j}+\oG_{[kj]i}\nonumber\\&=&
 \oG_{ijk}-C_{ijk}+C_{kij}+C_{kji}\,.
 \end{eqnarray}
The transformation law (\ref{cc1}) for this connection is a well known fact.
The invariance of  (\ref{LC1}) under arbitrary local
 $SO(1,3)$-transformations of the coframe field is clear from
 (\ref{LC-con}).

 We are  looking now for a most general connection $\G_{ijk}$
 which is linear in the first
 order derivatives of the coframe,
 i.e., for a generalization of (\ref{weiz2}) and (\ref{LC1}).
 Similarly to these special cases, the coefficients in the general
 linear combination of the first order derivatives have to be linear
 in the coframe components.
 In order to construct the most general coframe connection, we apply
 the well known fact: The difference of two arbitrary connections
 is a tensor.
 Thus a general coframe connection can be
 represented  as the Weitzenb\"{o}ck connection plus a
 tensor,
\begin{equation}\label{con}
\G_{ijk}=\oG_{ijk}+Y_{ijk}\,,
\end{equation}
 or, alternatively, as the Levi-Civita connection plus a
 tensor,
 \begin{equation}\label{conx}
\G_{ijk}=\LG_{ijk}+Z_{ijk}\,.
\end{equation}
The tensor $Y_{ijk}$ of (\ref{con}) has to be itself linear
 in the first order derivatives of the
 coframe field. Consequently it can be written as
\begin{equation}\label{cc14}
 Y_{ijk}=\chi_{ijk}{}^{mnl}\Big(\eta_{\a\b}\vt^\a{}_l\vt^\b{}_{[m,n]}\Big)=
 \chi_{ijk}{}^{mnl}\,C_{mnl}\,.
 \end{equation}
 The ``constitutive tensor''
 $ \chi_{ijk}{}^{mnl}$ can  involve only the
 components of the metric tensor and the Kronecker symbols.
 In view of  the symmetry relation
 $\chi_{ijk}{}^{mnl}=\chi_{ijk}{}^{[mn]l}$,
 it is enough to restrict  the  general expression to
 \begin{eqnarray}\label{chi}
\chi_{ijk}{}^{mnl}&=&\a_1\d_i^m\d_j^{n}\d_k^{l}+
g^{ml}(\a_2g_{ik}\d_j^n +\a_3g_{jk}\d_i^n+\b_1g_{ij}\d^n_k)+
 \nonumber\\
&&\quad\b_2\d^m_k\d^n_j\d^l_i+\b_3\d^m_k\d^n_i\d^l_j\,.
\end{eqnarray}
Substituting into (\ref{con}) we have
\begin{eqnarray}\label{g-con1}
 \G_{ijk}&=&\oG_{ijk}+\a_1C_{ijk}+\a_2g_{ik}C_j+\a_3g_{jk}C_i+
\nonumber\\
 &&\qquad\quad
 \b_1g_{ij}C_k+\b_2C_{kji}+\b_3C_{kij}\,,
 \end{eqnarray}
or, equivalently,
\begin{eqnarray}\label{g-con}
 \G_{ij}{}^k&=&\oG_{ij}{}^k+\a_1C_{ij}{}^k+\a_2C_j\d_i^k+
 \a_3C_i\d^k_j+\nonumber\\
 &&g^{kn}\left(\b_1g_{ij}C_{n}+\b_2g_{im}C_{nj}{}^m+
 \b_3g_{jm}C_{ni}{}^m\right)\,.
 \end{eqnarray}
The $\a$-terms in (\ref{g-con1}) are defined already
 on a linear connection manifold without a metric, the $\b$-terms require
  a metric structure on a manifold.
 We give an equivalent form of  the family (\ref{g-con})
via the Levi-Civita connection,
 \begin{eqnarray}\label{g-con2}
 \G_{ijk}&=&\LG_{ijk}+(\a_1+1)C_{ijk}+\a_2g_{ik}C_j+\a_3g_{jk}C_i+
\nonumber\\
 &&\qquad\quad
 \b_1g_{ij}C_k+(\b_2-1)C_{kji}+(\b_3-1)C_{kij}\,,
 \end{eqnarray}
 Hence, in contrast to Riemannian geometry with a unique Levi-Civita
 connection, we have, on the coframe manifold, an infinity
  6-parametric family of
connections constructed from the coframe components only.
\subsection{Torsion of the coframe connections}
On a  linear connection manifold, an asymmetric connection is
 characterized by its {\it torsion tensor} which is defined as
 \begin{equation}\label{tor}
T_{ij}{}^k=\G_{[ij]}{}^k\,.
\end{equation}
 This skew-symmetric tensor ($T_{ij}{}^k=-T_{ji}{}^k$) does not depend
 on the metric structure. For   the exterior form representation,
see \cite{Hehl:1994ue}---\cite{Itin:2003jp}.

For the  Weitzenb\"{o}ck connection, the torsion is given by
 $\oT_{ij}{}^k=C_{ij}{}^k$. Due to (\ref{weiz2}), this tensor is zero only
 for  closed coframe 1-forms which satisfy $d\vt^\a=0$.
The Levi-Civita connection is symmetric in the  down indices,
 thus its  torsion is zero.

 The torsion tensor of the general connection (\ref{g-con}) is given by
 \begin{eqnarray}\label{g-tor}
T_{ijk}&=&(\a_1+1)C_{ijk}+
\frac 12(\a_2-\a_3)(g_{ik}C_j-g_{jk}C_i)+\nonumber\\
 &&\frac 12(\b_2-\b_3)(C_{kji}-C_{kij})\,.
\end{eqnarray}
 Consequently, the relations
\begin{equation}\label{tor-free}
 \a_1=-1\,,\qquad \a_2=\a_3\,,\qquad \b_2=\b_3\,
\end{equation}
 extract a 3-parametric sub-family of identically
 symmetric (torsion-free) connections.

Geometrically, the non-zero torsion means a non-trivial parallel
 transport even on a flat manifold.
The field models on a manifold with a connection of a
 non-zero torsion can include additional terms for interaction with torsion.
 The physical effects of a non-zero torsion was studied intensively, see the
 reports \cite{Hehl:1976kj}---\cite{Hammond:2002rm} and the references
 given therein. Probably the most important result of these numerous studies
 is the fact that non-zero torsion does not contradict the standard
 physical paradigm.

\subsection{Non-metricity of the coframe connections}
Another algebraic characteristic of an asymmetric connection can be given
 on a manifold endowed with a metric.
 The {\it non-metricity tensor}, $Q_{kij}=Q_{kji}$, is defined
   by the  covariant
 derivative of the metric tensor,
\begin{equation}\label{nonmetr}
Q_{kij}=-g_{ij;k}=
 -g_{ij,k}+\G_{ikj}+\G_{jki}\,.
\end{equation}
where $\G_{ijk}=\G_{ij}{}^mg_{mk}$.

For the  Weitzenb\"{o}ck and the Levi-Civita connections,
 the non-metricity tensor is zero.
For the general connection (\ref{g-con}), it is given by
 \begin{eqnarray}\label{g-nonmetr}
Q_{kij}&=&(\a_1+\b_2)(C_{ikj}+C_{jki})+2\a_2g_{ij}C_k+
\nonumber\\
 &&(\a_3+\b_1)(g_{ik}C_j+g_{jk}C_i) \,.
\end{eqnarray}
Hence, the  connection (\ref{con}) is metric-compatible (has an identically
 zero tensor of non-metricity) if
  \begin{equation}\label{m-comp}
 \a_1=-\b_2\,, \qquad \a_2=0\,, \qquad \a_3=-\b_1\,.
 \end{equation}

  Two tensors, the torsion an the non-metricity,  characterize the
 affine connection uniquely \cite{Schou}.
 In fact, we can decompose the general affine
 connection (\ref{g-con}) as
\begin{eqnarray}\label{LC-decomp}
\G_{ijk}&=&\LG_{ijk}+(T_{ijk}+T_{kij}-T_{jki})+
 \frac 12 (Q_{ijk}-Q_{kij}+Q_{jki})\,.
\end{eqnarray}
Consequently, a manifold endowed with an asymmetric connection
 can be treated  as a Riemannian geometry with two additional tensors of
 torsion and non-metricity.
 Geometrically, a non-trivial non-metricity tensor means change of the
 lengths and angles when a parallel transport of vectors over a closed
 curve is applied. Such violation of rotational and Lorentz invariance
 bring us rather far
 from the standard physical paradigm.
 So, from the physical point of view,
 the metric-compatible constrains (\ref{m-comp}) are better
 motivated then the  torsion-free constrains (\ref{tor-free}).

When the requirements (\ref{tor-free},\ref{m-comp}) are
 considered together,
 we come  to a  unique torsion-free and metric-compatible
coframe connection given by the parameters
\begin{equation}\label{LC-par}
 \a_1=-1\,,\qquad \b_2=\b_3=1\,, \qquad \a_2=\a_3=\b_1=0\,.
 \end{equation}
 Certainly, it is no more than the ordinary Levi-Civita connection
 (\ref{LC-con}).

\subsection{The Riemannian curvature  of the coframe connections}
 Although the Riemannian curvature tensor  is a classical subject
 of differential geometry,
 in the case of an asymmetric connection, slightly different
 notations are in use. We accept  the agreements used in
 metric-affine gravity \cite{Hehl:1994ue}.

 The  Riemannian curvature tensor of a general asymmetric connection
 is defined  as
  \begin{equation}\label{CL6}
 R_{imn}{}^j=\G_{im}{}^j{}_{,n}-\G_{in}{}^j{}_{,m}+\G_{in}{}^k\,\G_{km}{}^j
 -\G_{im}{}^k\,\G_{kn}{}^j\,.
 \end{equation}
 We are interested in the Riemannian curvature of the coframe
 connection (\ref{g-con}).
 Due to (\ref{CL6}), $R_{imn}{}^j$  involves the parameters
 $\a_i,\b_i$ in linear and quadratic combinations.
 It  vanishes for zero values
 of the parameters, i.e., for the Weitzenb\"{o}ck
connection.
 Moreover, the Weitzenb\"{o}ck
connection $\oG_{ij}{}^k$ is a unique coframe connection of
 an identically zero curvature, for a proof, see  \cite{itin-con}.

 For classical fields  applications, we are interested in the scalar
 curvature, i.e., in the Hilbert-Einstein Lagrangian,
  \begin{equation}\label{HEL}
 \L=R\sqrt{-g}=g^{in}R_{ijn}{}^j\sqrt{-g}\,.
 \end{equation}
 When (\ref{g-con}) is substituted here, we obtain  \cite{itin-con}
  \begin{equation}\label{HE-tel}
 R\sqrt{-g}=C_{ijk}H^{ijk}\sqrt{-g}+{\rm \,\, total \,\,\, derivative}\,,
 \end{equation}
 where
 \begin{equation}\label{H-def0}
 H^{ijk}=\rho_1\,C^{ijk}+
3\rho_2C^{[ijk]}+\rho_3\Big(C^{ijk}-2\,C^{im}{}_mg^{jk}\Big)\,.
 \end{equation}
 The dimensionless coefficients $\rho_i$ are quadratic polynomials in the
 constants $\a_i,\b_i$.

 The quadratic part of the (\ref{HE-tel}) is well known from
 the gravity fields models based on the coframe fields \cite{Hayash:1979},
 \cite{Muench:1998ay},  \cite{Itin:2001bp}.
 Up to a change of parameters, it is the most general scalar expression
 quadratic in the first order derivatives of the coframe field.
 Usually, this expression is treated as an arbitrary linear combination
 of the squares of the Weitzenb\"{o}ck torsion.
 Note certain outputs of our coframe
 connections approach:
 \begin{itemize}
 \item[(i)] The quadratic coframe Lagrangian is regarded as a standard
 Einstein-Hilbert Lagrangian calculated on a general asymmetric
 coframe connection.
 \item[(ii)]  The six-parametric connection (\ref{g-con})
 is involved
 in the Lagrangian (\ref{HE-tel}) only via three independent combinations.
 Thus three additional requirements (for instance, the compatibility
 to the metric tensor) can be applied  without changing the physical
 content of the model. \end{itemize}

 \section{Gauge invariant connections}
\subsection{Gauge transformation}
 Let us examine  now the behavior of the geometrical structure
 (\ref{str2}) under local transformations of the coframe field
\begin{equation}\label{L-tr0}
\vt^\a\mapsto L^\a{}_\b(x)\,\vt^\b\,.
\end{equation}
The metric tensor itself is invariant under local Lorentz transformations,
thus, for every $x$, the matrix $ L^\a{}_\b(x) \in SO(1,3)$.
 Thus we are looking for a subclass of
 the connections (\ref{g-con1}) that invariant under pseudo-orthonormal
transformations  (\ref{L-tr0}).
 Since the coframe field appears in (\ref{str2}) only implicitly,
 (\ref{L-tr0}) is a type of  gauge transformation.
 Certainly, the Levi-Civita connection is in the desired subclass.
 Consequently, our
invariance condition does not bring us too far from the standard
Riemannian geometry. Moreover, with this requirement, we can
have only small additions also on the physical field level,  see
Sect 5.

In a coordinate basis,   the coframe components change under
(\ref{L-tr0}) as
 \begin{equation}\label{L-tr1}
 \vt^\a{}_i\mapsto L^\a{}_\b\,\vt^\b{}_i\,,\qquad
 e_\a{}^i\mapsto (L^{-1})^\b{}_\a\, e_\b{}^i\,.
\end{equation}
In this paper, we will deal only with the  infinitesimal version of the
transformations (\ref{L-tr0}). Consequently, we restrict to
\begin{equation}\label{inf-L}
L^\a{}_\b=\d^\a_\b+X^\a{}_\b\,.
\end{equation}
Correspondingly, the matrix $X_{\a\b}=X^\mu{}_\b \eta_{\mu\a}$
 is antisymmetric.
 Under (\ref{inf-L}), the components of the basis fields change as
 \begin{equation}\label{L-tr2}
 \vt^\a{}_i\mapsto \vt^\a{}_i+X^\a{}_\b\,\vt^\b{}_i\,,\qquad
 e_\a{}^i\mapsto e_\a{}^i -X^\b{}_\a\, e_\b{}^i\,.
\end{equation}
It should be noted that  our analyses is restricted
 to the linear approximation of the transformation matrix $L^\a{}_\b$.
 In particular, we neglect with the second order term in the transformation
 of the metric tensor and assume it  invariant under (\ref{L-tr2}).

 \subsection{Invariance of the general connections}
We are examining now, under what conditions, the connections
$\G_{ij}{}^k(\vt^\a)$
 are invariant under the transformations (\ref{L-tr2}).
 The  Weitzenb\"{o}ck connection changes as
 \begin{equation}\label{W-change}
 \delta \oG_{ijk}=\vt^\a{}_k\vt^\b{}_iX_{\a\b,j}\,.
 \end{equation}
 This connection is invariant, $ \delta \oG_{ijk}=0$,  only if $X_{\a\b,a}=0$,
 i.e., only for global (rigid) transformations.
Recall, that the torsion of this connection is a building block of our
construction.
 Denoting its variation by $K_{ijk}$, we have from (\ref{W-change})
\begin{equation}\label{C-change}
 K_{ijk}=\delta C_{ijk}=\frac 12 \, \vt^\a{}_k\Big(X_{\a\b,j}\vt^\b{}_i-
 X_{\a\b,i}\vt^\b{}_j\Big)\,.
 \end{equation}
Correspondingly,
\begin{equation}\label{C-change-x}
 K_{mj}{}^m=g^{ik}\delta C_{ijk}=\delta(g^{ik} C_{ijk})=\delta C_i\,.
\end{equation}

For  variation of the general connection, we use the form
(\ref{g-con2}), in which the invariant part (the Levi-Civita connection) is
 extracted explicitly.
Thus the invariance condition , $\d(\G_{ijk})=0$, is expressed as
  \begin{eqnarray}\label{g-con2x}
 &&(\a_1+1)K_{ijk}+\a_2g_{ik}K_{mj}{}^m+
 \a_3g_{jk}K_{im}{}^m+\nonumber\\
 &&\qquad \qquad \b_1g_{ij}K_{km}{}^m+(\b_2-1)K_{kji}+(\b_3-1)K_{kij}=0\,.
 \end{eqnarray}
 Let us examine now how this equation can be satisfied: \vspace{3 pt}

 \noindent (i) {\it Local transformations.} Certainly,
 (\ref{g-con2x}) holds when all its coefficients are zero, i.e., the
 connection is of Levi-Civita. The quantity
 $K_{ijk}$ is arbitrary in this case, thus also the transformation matrix
 $X_{\a\b}$ is non-restricted.
 Thus arbitrary $SO(1,3)$ transformations of the coframe
 are acceptable as it has to be in Riemannian geometry. \vspace{3 pt}

 \noindent (ii) {\it Rigid transformations.} Another type of solution
 to (\ref{g-con2x}) emerges by requiring $K_{ijk}=0$.
 In this case we have only rigid transformations of the coframe,
 which are certainly admissible for all coframe connections. \vspace{3 pt}

  \noindent (iii) {\it Gauge transformations.} We are looking  now for a
 nontrivial solution to  (\ref{g-con2x}), such that the
 the elements of the matrix $X_{\a\b}$ are dynamical,
 i.e., satisfy a well posed system of partial differential equations.
 \subsection{Gauge invariant connections}
 A first consequence of (\ref{g-con2x}) is its trace, i.e.,
 the contraction in two indices. Since the left hand side of the equation is
asymmetric,
 we have here three different traces all of the same form:
\begin{equation}\label{trace}
\lambda_i K^m{}_{im}=0\,.
\end{equation}
These conditions  are necessary for invariance of the connection.
Here, the coefficients $\lambda_i$ are equal to three different linear
combinations of the parameters $\a_i\,,\b_i$. In
particular, all $\lambda_i$ are zero for the Levi-Civita
connection. We will consider, however, an alternative solution
with at least one nonzero parameter $\lambda_i$. So we derive that
 the variations of the
coframe field  have to satisfy the equation
\begin{equation}\label{trace1}
K^m{}_{im}=0\,.
\end{equation}

Substituting  it into (\ref{g-con2x}) we obtain
\begin{equation}\label{trace2}
(\a_1+1)K_{ijk}+(\b_2-1)K_{kji}+(\b_3-1)K_{kij}=0,.
\end{equation}
The completely antisymmetric combination of this equation yields
\begin{equation}\label{trace3}
(\a_1-\b_2+\b_3+1)K_{[ijk]}=0\,.
\end{equation}
Again, for the Levi-Civita connection, the coefficient of this equation is
zero. In the alternative case of a nonzero coefficient, the
variations of the coframe field has to satisfy the equation
\begin{equation}\label{trace4}
K_{[ijk]}=0\,.
\end{equation}
 Certainly this result holds only for $\a_1+1\ne \b_2-\b_3$, otherwise we
 do not have the condition (\ref{trace3}) at all.
 We neglect with  this special case because it gives an undetermined
system of conditions on $X_{\a\b}$.

When  (\ref{trace1}) and (\ref{trace4}) are substituted into
 (\ref{g-con2x}) we remain with
\begin{equation}\label{trace5}
(\a_1-\b_3+2)K_{ijk}+(2-\b_2-\b_3)K_{kji}=0\,.
\end{equation}
We have to restrict now  the  coefficients, otherwise we obtain $K_{ijk}=0$,
 i.e., only the rigid transformations.
Hence,
 \begin{equation}
\a_1-\b_3+2=0\,,\qquad 2-\b_2-\b_3=0\,,
\end{equation}
or, equivalently,
 \begin{equation}\label{par1}
\b_2=-\a_1\,, \qquad \b_3=2+\a_1\,.
\end{equation}
Consequently, we have derived a family of coframe connections
\begin{eqnarray}\label{g-con2g}
 \G_{ijk}&=&\oG_{ijk}+2C_{ikj}-3\a_1C_{[ijk]}+
 \nonumber\\
&&\qquad
 \a_2g_{ik}C_k+\a_3g_{jk}C_i+\b_1g_{ij}C_k\,,
 \end{eqnarray}
which are invariant under local restricted variations of the
 coframe field.
 In this expression, the coefficients $\a_1\,,\a_2\,,\a_3\,,b_1$
 can be taken almost arbitrary. Certainly, some exceptional values,
 for instance $\a_1=-1$,  are forbidden.
The torsion of the connection (\ref{g-con2g}) is
  \begin{eqnarray}\label{m-tor}
T_{ijk}&=&-3(1+\a_1)C_{[ijk]}-(\a_2-\a_3)C_{[i}g_{j]k}\,,
\end{eqnarray}
while the non-metricity tensor is
 \begin{eqnarray}\label{m-nonmetr}
Q_{kij}&=&2\a_2g_{ij}C_k+2(\a_3+\b_1)C_{(i}g_{j)k} \,.
\end{eqnarray}
\subsection{Gauge invariant metric compatible connections}
Starting with a coframe field on a manifold $\M$ we derived a most
general six-parametric family of connections which are linear in the first
order derivatives of the coframe components. In this family, we identified,
 three subclasses of torsion-free,
metric-compatible, and gauge invariant connections. Let us
compare the conditions determined these subclasses. Since  the
gauge invariant condition (\ref{par1}) contradicts to
(\ref{tor-free}), the gauge invariant connection cannot be
torsion-free.  As for the conditions of metric-compatibility
(\ref{m-comp}), they are correlated  (even partially overlap) with
(\ref{par1}). Consequently, we can introduce a subclass of {\it gauge
invariant metric compatible connections} with the parameters
\begin{equation}\label{join-par}
\a_2=0\,,\qquad \b_1=-\a_3\,,\qquad \b_2=-\a_1\,,\qquad
\b_3=2+\a_1\,.
\end{equation}
Two parameters $\a_1$ and $\a_3$ remain free. The corresponded
connection is
\begin{eqnarray}\label{join-con1}
 \G_{ijk}&=&\oG_{ijk}+2C_{ikj}-3\a_1C_{[ijk]}+
\a_3(g_{jk}C_i-g_{ij}C_k)\,.
 \end{eqnarray}
or, equivalently,
\begin{eqnarray}\label{join-con2}
 \G_{ijk}&=&\LG_{ijk}-3(\a_1+1)C_{[ijk]}+
\a_3(g_{jk}C_i-g_{ij}C_k)\,.
 \end{eqnarray}
 Although two parameters $\a_1$ and $\a_3$ remain free, the torsion of
 (\ref{join-con1})
 \begin{eqnarray}\label{join-tor}
T_{ijk}&=&-3(1+\a_1)C_{[ijk]}+\a_3C_{[i}g_{j]k}\,.
\end{eqnarray}
 cannot be zero. Recall that some values of the parameters, in particular,
 $\a_1\ne 1$, are forbidden.

 We summarize the relations between different coframe connections
 in Fig. 1.

\begin{figure}[t]
\begin{minipage}{16cm}
 \centering
 \epsfxsize=2.25in
\includegraphics[width=12cm]{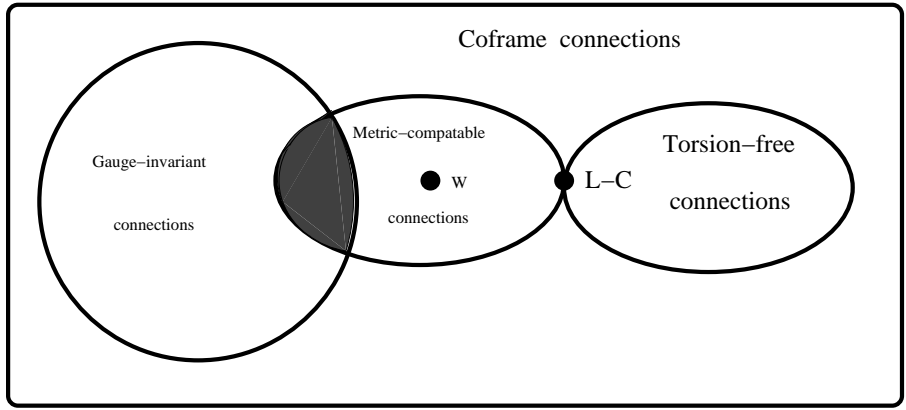}
 \end{minipage}
\caption{
 The different connections are depicted by the interior
 points of the rectangle.
 The bold points denote the Weitzenb\"{o}ck and the Levi-Civita
 connections. The gray area represents the gauge-invariant
metric-compatible connections.}
\end{figure}


The standard Levi-Civita connection is invariant under arbitrary
non-restricted variations of the coframe. This connection is a
basis construction of the Riemannian geometry and consequently
 of the Einstein gravity theory.

 We have derived an alternative  family of connections which
 are invariant under restricted local variations of the
 coframe field.  The variations themselves have to satisfy
 the equations
\begin{eqnarray}\label{PreMax}
K_{[abc]}=0\,, \qquad K^m{}_{am}=0\,.
\end{eqnarray}
Due to (\ref{C-change}) this is a system of eight first order
partial differential equations for six enters of the
antisymmetric matrix $X_{\a\b}$. This is very similar to the
standard Maxwell system for the electromagnetic field strength.
We will examine the system (\ref{PreMax}) in section 5. But first
we will give a gravity model corresponding to these gauge
connections.
\section{Coframe field models}
\subsection{Action and field equation}
 The Lagrangian of a gauge invariant gravity model has to respect
 the gauge transformations (\ref{L-tr2})  restricted by (\ref{PreMax}).
 Starting with invariant fields of metric and connection, such
 Lagrangian can be constructed straightforwardly.
 Namely, we can take the standard Einstein-Hilbert Lagrangian calculated
 on our gauge  invariant connections instead of the Levi-Civita connections
  \begin{equation}\label{SL1}
 \A=\int R\Big(\G_{ab}{}^c\Big)\sqrt{-g}\,d^4x\,.
 \end{equation}
 Moreover, since our connection is asymmetric, we can also consider the terms
 quadratic in torsion and non-metricity as acceptable additions to the
 Lagrangians. All such terms are of the same physical dimensions so
 they have to be involved with free dimensionless coefficients.
 This way we come to a rather complicated Lagrangian of
 MAG \cite{Hehl:1994ue}.

 In our case, however, all the geometric
 quantities are constructed only
 from the coframe components and their first order derivatives.
 This fact simplifies very much the most general Lagrangian
 we are seeking for.  Observe that the Einstein-Hilbert Lagrangian (\ref{SL1})
 can be
 rewritten as a linear combination of the first order derivatives of
 the coframe components plus a total derivative term.
 Also the  torsion and non-metricity tensors are linear in
 the first order derivatives of the coframe. Consequently, the most general
 Lagrangian can be given by a linear combination of terms which are
 quadratic in the first order derivatives of the coframe.

A most general coframe Lagrangian
 can be  straightforwardly constructed from the  exterior derivative
 of the coframe components $\vt^\a{}_{[i,j]}$,
\begin{equation}\label{C-der}
d\vt^\a=\vt^\a{}_{[i,j]}dx^i\wedge dx^j\,.
\end{equation}
Equivalently, we can use the components $\vt^\a{}_{[i,j]}$
directly to deal with the tensorial expression
\begin{equation}\label{C-def1}
 C_{ij}{}^k=e_\a{}^k\vt^\a{}_{[i,j]}\,,\qquad
C_{ijk}=\eta_{\a\b}\vt^\a{}_k\vt^\b{}_{[i,j]}\,.
\end{equation}
Now it is enough to consider the possible quadratic combinations
of $d\vt^\a$ or, equivalently, of $C_{ij}{}^k$.
 In this way,  a general quadratic coframe action
takes the known form \cite{Muench:1998ay}, \cite{Itin:2001bp}
\begin{equation}\label{cof-lag0}
{}^{\tt (cof)}\A= \frac {\kappa} 2\int\rho_1 L_1+\rho_2\,L_2+\rho_3 \,L_3\,,
\end{equation}
where $\kappa$ is a coupling constant of dimension $1/\ell^2$,
 $\rho_i$ are free dimensionless parameters, while the
 ``elementary'' Lagrangian densities are defined as
\brr\label{cof-lag1}
L_1 &=&d\vt^\a \wedge *d\vt_\a,\\
\label{cof-lag2} L_2 &=&
(d\vt_\a \wedge \vt^\a ) \wedge*(d\vt_\b\wedge\vt^\b), \\
\label{cof-lag3} L_3 &=& (d\vt_\a \wedge\vt^\b ) \wedge
*(d\vt_\b \wedge \vt^\a )\,.
  \err
 The action
(\ref{cof-lag0}) can be rewritten in a compact form
\begin{equation}\label{cof-lag4}
{}^{\tt (cof)}\A=\frac {\kappa} {2} \int d\vt_\a\wedge *H^\a=
 \frac{\kappa}2\int C_{ijk} H^{ijk}\,\sqrt{-g}\,dx^4\,,
 \end{equation}
 where
  \begin{equation}\label{H-def}
 H^{ijk}=\rho_1\,C^{ijk}+
3\rho_2C^{[ijk]}+\rho_3\Big(C^{ijk}-2\,C^{im}{}_mg^{jk}\Big)\,.
 \end{equation}
 The corresponding 1-form is
  \begin{equation}\label{H-def1}
 H^\a=\frac 12 \, H_{ijk}\vt^\a{}_k dx^i\wedge dx^j\,.
  \end{equation}

To make physics on the coframe background we have to involve  a
Lagrangian of a matter field $\psi$. We do not specify the
tensorial or spinorial content of this field. Consequently,
\begin{equation}\label{mat-lag}
{}^{\tt (mat)}\A=\int L(\psi, \nabla_\Gamma\psi, g)\,,
\end{equation}
 where the integrand is a 4-form density.
The covariant derivative in (\ref{mat-lag}) has to be
 taken with respect to the same
connection $\Gamma$ which is involved in the coframe Lagrangian.
 Consequently, also ${}^{(m)}\A$ involves the  coframe components.
Apply now the variation of the  total Lagrangian
 \begin{equation}\label{mat-lag1}
\A={}^{\tt (cof)}A+{}^{\tt (mat)}A
\end{equation}
 with respect to the coframe field.  We come to the coframe field equation
 which has a compact form in exterior calculus notation \cite{Itin:2001bp}
  \begin{equation}\label{cof-feq}
d*H^\a={}^{\tt (cof)}\Sigma^\a+{}^{\tt (mat)}\Sigma^\a\,,
\end{equation}
where the right hand side involves the energy-momentum currents (3-form)
 of the
 coframe and the matter fields. These quantities are related to the
 energy-momentum tensors $T^{ij}$ in a regular way
\begin{equation}
 \Sigma^\a=(1/6)\,T^{ij}\,\vt^\a{}_i\,\epsilon_{jklm}dx^k\wedge
 dx^l\wedge dx^m\,.
 \end{equation}

 The standard Einstein-Hilbert action of GR is not
 completely distinct from the
 coframe action (\ref{cof-lag0}).
 Indeed, the term $R\sqrt{-g}$ can be
rewritten as a total derivative plus squares of first order derivatives
of the metric. Let us substitute
$g_{ij}=\eta_{\a\b}\vt^\a{}_i\vt^\b{}_j$ and neglect with the
total derivative. We remain with a scalar expression which is
quadratic in the coframe components.
 Since all such type expressions are encoded in
(\ref{cof-lag0}),  there is a coframe Lagrangian
with a special set of parameters   is equivalent  (up to a total
derivative) to the standard Hilbert-Einstein Lagrangian. On the
level of the field equations, we have here no more than the
standard GR reformulated in the coframe variables.
The explicit calculations \cite{Muench:1998ay} give  the
 special (Einstein) choice of parameters
  \begin{equation}\label{E-choice}
\rho_1=0\,,\qquad \rho_2=- 1/2 \,,\qquad \rho_3=1
\end{equation}
Consequently, up to a total derivative, the Hilbert-Einstein action takes the
form
\begin{equation}\label{E-choice1}
\int R\sqrt{-g}\,{\rm vol}=\int \Big(-\frac 12 \,L_2+\,L_3\Big)\,.
\end{equation}
Note that although the energy-momentum tensor of the coframe field is
 defined also in this case, it cannot be identified as an
 energy-momentum tensor of gravity. Such identification quite often appears in
 literature, because one does not take into account a hidden symmetry
 of the coframe action. For (\ref{E-choice1}), the action  (\ref{cof-lag0})
 is local Lorentz invariant.
 This symmetry, however, is not preserved for the coframe
 energy-momentum tensor expression.

 \subsection{Gauge invariant coframe Lagrangian}

 Let us  examine now the behavior of the general coframe action
 (\ref{cof-lag0})
 under the local Lorentz transformations (\ref{L-tr2}).
 Using (\ref{E-choice1})  we  rewrite (up to a total derivative)
\begin{equation}\label{act-trans}
\int {}^{\tt {(cof)}}L=\int \Big[\rho_1 L_1+(\rho_2+\frac 12 \rho_3)\,L_2
 \Big]+\rho_3\int R\sqrt{-g}\,{\rm vol}\,.
\end{equation}
With respect to the transformations (\ref{L-tr2}),   this
expression changes as
\begin{eqnarray}\label{var-act}
 \d\int {}^{\tt {(cof)}}L&=&\int \Big[\rho_1\d L_1+(\rho_2+\frac 12 \rho_3)\d L_2
  \Big]\nonumber\\
 &=&\int C_{ijk}\Big[2\rho_1 K^{ijk}+(2\rho_2+\rho_3)K^{[ijk]}\Big]
 \sqrt{-g}\,{\rm vol}\,.
 \end{eqnarray}
 Since the variations $K^{ijk}$ are independent, the action is
 invariant  if and only if
 the following invariance condition holds
 \begin{equation}\label{requ0}
2\rho_1 K^{ijk}+(2\rho_2-\rho_3) K^{[ijk]}=0\,.
\end{equation}
This algebraic equation has to be satisfied identically (for arbitrary coframe fields), thus we have
three solutions to (\ref{requ0}) of basically  different types:

 \hspace{0.5 cm}

\noindent  {\it (i) "The teleparallel equivalent of GR"}

\noindent  A first solution to  (\ref{requ0}) can be taken as
\begin{equation}\label{requ2}
  \rho_1=0\,,\qquad 2\rho_2+\rho_3=0\,.
\end{equation}
 Since the tensor $K^{ijk}$ is arbitrary now, the
coframe Lagrangian with (\ref{requ2}) is invariant under arbitrary
Lorentz transformations. Consequently,   six degrees of freedom of the
coframe field are unphysical while the rest ten degrees are
equivalent to the metric tensor. The underlining geometry for
such coframe model is precisely Riemannian and not teleparallel as
it is claimed sometimes. The fact that the coframe Lagrangian is
expressed by the torsion of the  Weitzenb\"{o}ck connection does
not mean a lot, specially in view of the formula (\ref{LC-con}),
which express the Levi-Civita connection by the Weitzenb\"{o}ck
one.

The coframe reformulation of GR can serve as a useful technical
tool, for instance  for search of new solutions in GR
\cite{Obukhov:2002tm} or for its Hamiltonian formulation
\cite{Blagojevic:2000pi}. It cannot change however the main features of GR. In
particular, the so-called teleparallel energy-momentum tensors of
gravitational field are no more than the pseudo-tensors of the
standard GR.

 \hspace{0.5 cm}

\noindent {\it (ii) The teleparallel gravity model.}

 \noindent Another solution to  (\ref{requ0}) is
\begin{equation}\label{requ1}
  K^{ijk}=0\,,\qquad \rho_1\ne 0\,.
\end{equation}
Substituting into (\ref{C-change}) we obtain $X_{\a\b,i}=0$, thus
only global (rigid) Lorentz transformations of the coframe field
are admissible.  Consequently, if a coframe is given at a point, it is also
 given on the whole manifold. Such teleparallel geometry is described
by  the  Weitzenb\"{o}ck connection.
 Consequently, precisely the model (\ref{requ1})
 (and only this model) has to be
refereed to as a teleparallel gravity model. Indeed only the
Lagrangians with $\rho_1\ne 0$ do not have any local non-trivial
group  of transformations. This is in a complete correspondence
to the flat teleparallel geometry.

With a non-zero parameter $\rho_1$, the coframe field equation
 (\ref{cof-feq}) has  spherical symmetric solutions
 of the type \cite{Itin:1999wi}
\begin{equation}\label{non-Sc}
\vt^\a= \left(\frac r{r_o}\right)^Adx^\a\,,
\end{equation}
 where $A$ is a function of the parameters $\rho_i$.
 The corresponding metric does not have the
 Newtonian behavior at infinity. Consequently, the teleparallel
 model cannot serve for description of $4D$ gravitational field.

 \hspace{0.5 cm}

\noindent (iii) {\it A model alternative to GR}

\noindent The last solution to (\ref{requ0}) is
\begin{equation}\label{requ3}
  K^{[ijk]}=0\qquad \rho_1=0\,,\qquad 2\rho_2-\rho_3\ne 0\,,
\end{equation}
The first equation here is a first order partial differential
equation for the matrix $X_{\a\b}$. Thus, in contrast to the previous cases,
 we are dealing now with {\it
dynamical local Lorentz transformations}.

For the coframe models with a zero parameter $\rho_1$, the field
equation (\ref{cof-feq}) has a unique static spherical-symmetric
solution of a "diagonal form" \cite{Itin:1999wi} :
\begin{equation}\label{Sch}
\vt^0=\frac{1-m/2\rho}{1+m/2\rho}\,dx^0\,,\qquad \vt^i=
\left(1+\frac{m}{2\rho}\right)^2\,dx^i\,,
\end{equation}
where $i=1,2,3$. This coframe corresponds to the Schwarzschild
metric in the isotropic coordinates.

Another justification for the condition $\rho_1=0$ comes from consideration
of the first order approximation to the coframe field model
\cite{Itin:2004ig}. In this case, the coframe variable is reduced
to a sum of symmetric and antisymmetric matrices. It means that,
in linear approximation,
 we can treat the coframe field as a system
of two independent fields. It is natural to require all the
field-theoretic constructions, i.e. the action, the field equation
and the energy-momentum tensor, to accept the same separation to
two independent expression. A remarkable fact that such
separation (free field limit) appears if and only if $\rho_1=0$.

 Consequently we have derived the equation
\begin{equation}\label{requ4}
  K^{[ijk]}=0\,,
\end{equation}
 as an invariance condition for a viable gravity field model.
 This is in an addition to the pure geometrical consideration given in  the
 previous section.
 \subsection{Gauge invariant matter Lagrangian}
We turn now to the second constrain $K^m{}_{im}=0$.
Distinctly, it comes from the symmetry a generic matter Lagrangian.
 When the matter field $\psi$ is minimally
coupled to gravity, its Lagrangian involves the covariant
derivatives taken with respect to some connection
\begin{equation}\label{mat1-lag}
{}^{(m)}L=L(\psi, \nabla_\Gamma\psi, g)\,.
\end{equation}
Since the matter fields themselves are invariant under coframe
transformations, also the connection $\Gamma$ has to be invariant.
In other words,
\begin{equation}
 \d({}^{(m)}L)=0\quad {\rm  yields}\quad  \d\G_{ijk}=0\,.
\end{equation}
 In the case of
 a variety of connections the following problem emerges \cite{gyros}:
 To what connection
the matter field is really coupled?  A natural answer can be
proposed: The proper connection is this one which is already
involved in the gravity sector of the model. In other words: The
symmetries of the gravity and the matter sectors have to be
conformed.

We continue now with our alternative model
(\ref{requ3}). Since the gravity sector is already
restricted with the requirement
$K_{[ijk]}=0$, it is natural to
require the matter sector to respect this condition. This way we
come to the family of connections (\ref{g-con2g}) which are
invariant under the local Lorentz transformations satisfied
 \begin{equation}
K^m{}_{im}=0\,.
\end{equation}
Consequently, the equations (\ref{PreMax}) emerge  as
the invariance conditions for the total action of a matter field
coupling minimally to gravity.

 \section{Maxwell-compatible connection}
 \subsection{Invariance conditions on the flat space}
Let us examine  now what physical meaning can be given to the
invariance conditions
\begin{eqnarray}\label{PreMax-x}
K_{[ijk]}=0\,, \qquad K^m{}_{im}=0\,.
\end{eqnarray}
Recall that the tensor $ K_{ijk}$ depends on the derivatives of the Lorentz
parameters $X_{\a\b}$ and on the components of the coframe field
\begin{equation}\label{KK-def}
 K_{ijk}=\frac 12 \, \vt^\a{}_k\Big(X_{\a\b,j}\vt^\b{}_i-
 X_{\a\b,i}\vt^\b{}_j\Big)\,.
 \end{equation}
 Thus, in fact, we have in (\ref{PreMax-x}), two first order
  partial differential equations
  for the entries of an  antisymmetric matrix $X_{\a\b}$.
Let us construct from this matrix  an antisymmetric  tensor $F_{ij}$
 \begin{equation}\label{F1-def}
 F_{ij}=X_{\mu\nu}\vt^\mu{}_i\vt^\nu{}_j\,,\qquad
X_{\mu\nu}=F_{ij}e_\mu{}^ie_\nu{}^j\,.
 \end{equation}
Substituting into (\ref{KK-def}), we derive
\begin{eqnarray}\label{KK1-def}
 K_{ijk}&=&F_{k[i,j]}-\frac 12 \, X_{\a\b}\Big[(\vt^\a{}_k\vt^\b{}_i)_{,j}-
 (\vt^\a{}_k\vt^\b{}_j)_{,i}\Big]\nonumber\\
 &=&F_{k[i,j]}-F_{km}C_{ij}{}^m-\frac 12 (F_{mi}\oG_{kj}{}^m-
 F_{mj}\oG_{ki}{}^m)\,.
 \end{eqnarray}
Consequently, the first equation from (\ref{PreMax-x}) takes the
form
\begin{eqnarray}\label{PreMax1-x}
F_{[ij,k]}=\frac 23
(C_{ij}{}^mF_{km}+C_{jk}{}^mF_{im}+C_{ki}{}^mF_{jm})\,,
\end{eqnarray}
while the second equation from (\ref{PreMax-x}) is rewritten as
\begin{eqnarray}\label{PreMax2-x}
F^i{}_{j,i}=-2F^i{}_mC_{ij}{}^m+
F_{kj}g^{ki}{}_{,i}+F_{mj}g^{ki}\oG_{ki}{}^m-F_{mi}g^{ki}\oG_{kj}{}^m\,.
\end{eqnarray}
Observe first  a significant approximation to
(\ref{PreMax1-x}---\ref{PreMax2-x}). If  the right hand sides in
 both equations are
neglected, the equations take the form of the ordinary Maxwell
equations for the electromagnetic field in vacuum ---
 \begin{equation}\label{Max-Lor}
 F_{[ij,k]}=0\,, \qquad F^i{}_{j,i}=0\,.
 \end{equation}
In the coframe models,  the gravity is modeled by a variable
coframe field, i.e., by nonzero values of the quantities
$\oG_{ij}{}^k$. Consequently, the right hand sides of
(\ref{PreMax1-x}---\ref{PreMax2-x}) can be viewed as curved space
additions, i.e., as the  gravitational corrections to the
electromagnetic field equations.
In the flat spacetime, when a suitable coordinate system is chosen,
 these corrections are identically equal to zero. Consequently,
 in the flat spacetime, the invariance conditions
 (\ref{PreMax-x}) take the form of  the  vacuum Maxwell
 system.

\subsection{Invariance conditions  on a  curved space}
On a curved manifold, the standard Maxwell equations are
formulated in a covariant form. Let us show that our
system
(\ref{PreMax1-x}---\ref{PreMax2-x}) is already covariant.
We rewrite (\ref{KK1-def}) as
\begin{eqnarray}\label{KK2-def}
 K_{ijk}=\frac 12(F_{ki,j}-F_{km}\oG_{ij}{}^m-
 F_{mi}\oG_{kj}{}^m)-
 \frac 12 (\,\,i\longleftrightarrow j\,\,)\,.
 \end{eqnarray}
Consequently,
\begin{eqnarray}\label{KK3-def}
 K_{ijk}=F_{k[i;j]}\,,
  \end{eqnarray}
where the covariant derivative (denoted by the semicolon)
is taken relative to the Weitzenb\"{o}ck connection.
 Consequently, the system
(\ref{PreMax1-x}---\ref{PreMax2-x}) takes the covariant form
\begin{equation}\label{Max-Rie}
 F_{[ij;k]}=0\,, \qquad F^i{}_{j;i}=0\,.
 \end{equation}
 These equations are literally the same as the
 electromagnetic sector field equations of the Maxwell-Einstein system.
 The crucial difference is  encoded in the type of the covariant derivative.
 In the Maxwell-Einstein system, the covariant derivative is taken relative
 to the Levi-Civita connection, while, in our case,
 the corresponding connection is of   Weitzenb\"{o}ck.
 Observe that, due to our approach, the Weitzenb\"{o}ck connection is rather
natural in (\ref{Max-Rie}). Indeed, since the electromagnetic-type 
 field describes
 the local change of the coframe field, it should itself be referred only
 to the global changes of the coframe. As we have shown, such
 global transformations correspond precisely to the teleparallel
 geometry with the   Weitzenb\"{o}ck connections.

\section{Gravity corrections to  the Coulomb-type field}
\subsection{Gravity-electromagnetic coupling}

 The coupling between the electromagnetic and the gravitational field
is an age-old problem. It is already related to the first observable
prediction of GR about the bending of light rays of stars in the
gravitational field of the Sun. The electromagnetic and gravitational
effects are of rather different orders of magnitude. However, the
increasing precision of modern experimental techniques   gives rise to
the hope that the appropriate form of the coupling can soon be
determined.

In particular, we have in this context two independent but closely
related problems:
\begin{itemize}
\item[(i)] Does the gravitational field of a charged massive source depend
  on the electric charge?
\item[(ii)] Does the electromagnetic field of a charged massive
  source depend on the mass?
\end{itemize}
From a certain philosophical point of view
 (``Everything has an influence on everything else'') the answer on both
 questions has to be positive.

In classical (non-relativistic) physics, the both answers
 are negative: The Newton force of attraction is independent on the
 charges, also the Coulomb force is not sensitive to the masses.

 \subsection{Gravity-electromagnetic coupling in GR}

In the framework of GR,  the coupling between the electromagnetic
field and gravity is  managed by the electromagnetic  action
itself
\begin{equation}\label{EM-act}
  S(g,F)=-\frac{\lambda_0}{4}\int F_{ij}F^{ij}\sqrt{-g}\,d^4x\,,
\end{equation}
where $\lambda_0$ is a coupling constant.
 When the actions of the gravitational and the
matter fields are added  to (\ref{EM-act}),  the variation with respect
 to the metric yields the
Einstein field equation (without cosmological constant)
\begin{equation}\label{Einst-eq1}
  R_{ij}-\frac 12 \, Rg_{ij}=\frac{8\pi G}{c^3}\Big( {}^{\tt
    (em)}T_{ij}+{}^{\tt (mat)}T_{ij}\Big)\,.
\end{equation}
Here the electromagnetic energy-momentum tensor ${}^{\tt(em)}T_{ij}$ and
 the matter energy-momentum tensor ${}^{\tt (mat)}T_{ij}$  are the
sources of the gravitational field.
The action (\ref{EM-act}) yields the electromagnetic field
equation of the form
 \begin{equation}\label{Max-eq1}
 F_{[ij;k]}=0\,, \qquad F^i_{j;i}=J_j\,,
 \end{equation}
 where the semicolon is used for the covariant derivative taken relative
 to the Levi-Civita connection.
 The dependence of the metric is encoded here  in the index raising procedure
 and in the covariant derivatives.

 The static spherical-symmetric solution to the Einstein-Maxwell system
 (\ref{Einst-eq1}, \ref{Max-eq1}) is given by the  Reissner-Nordstr\"om
 metric
 \begin{eqnarray}\label{RN-sol}
  &&ds^2=\big[1-\lambda(r)\big]dt^2- \big[1-\lambda(r)\big]^{-1}dr^2-
 r^2d\Omega^2,
\end{eqnarray}
 with
 \begin{eqnarray}\label{RN-sol1}
 &&\lambda(r)=\frac{2m}r-\frac {q^2}{r^2}\,,
\end{eqnarray}
 and the Coulomb potential
  \begin{equation}\label{Cou-sol}
 A_o=\frac Q r\,.
 \end{equation}
Here $m=GM/c^2$ and $q^2=GQ^2/(4\pi\varepsilon_0c^4)$ describes a
mass $M$ with an electric charge $Q$.
Consequently, in the Einstein-Maxwell system, the gravitational
 field depends on the charge of the source. The electromagnetic
 field of a pointwise source remains the same as in the flat spacetime, i.e.,
 it is independent on the mass of the charge.

 \subsection{Mass corrections to the Coulomb-type field}
 Let us look for a static spherically symmetric solution to our
electromagnetic-type equations (\ref{Max-Rie}).
 We start with a spherically symmetric solution in a vacuum coframe model
 (\ref{cof-lag0}) with the parameter
 $\rho_1=0$.
We will use the isotropic coordinates $\{x^\hi\,,\,
\hi=1,2,3\}$ with the isotropic radius
\begin{equation}
 \rho=\sqrt {\d_{\hi\hj}x^\hi
x^\hj}=\sqrt{x^2+y^2+z^2}\,.
\end{equation}

 Recall that we identify the gravity variable with the coframe field defined
 up to an infinitesimal  Lorentz transformation.
 It is equivalent to the metric field.
 Since the field equation (\ref{cof-feq}) involves free parameters
 $\rho_2$ and $\rho_3$, it is alternative  to GR.
 Although, the spherically symmetric gravity
 solution turns out to be the same. In particular, such solution
 can be taken in the form of
 a ``diagonal'' coframe  \cite{Itin:1999wi}
\begin{equation}\label{gr-anz}
\vt^0=\varphi(\rho)\,dx^0\,,\qquad \vt^{\hi}=
\psi(\rho)\,dx^{\hi}\,,
\end{equation}
where
\begin{equation}\label{gr-f-anz}
\varphi(\rho)=\frac{1-m/2\rho}{1+m/2\rho} \,,\qquad
\psi(\rho)=\left(1+\frac{m}{2\rho}\right)^2\,.
\end{equation}
The non-zero components of the corresponding metric tensor are
 \begin{equation}\label{gr-metr}
 g_{00}=-\varphi^2\,,\qquad g_{\hi \hi}=\psi^2\,,
 \end{equation}
 while the line element is
 \begin{equation}\label{line}
 ds^2=-\varphi^2c^2dt^2+\psi^2(dx^2+dy^2+dz^2)\,.
 \end{equation}
 Recall that (\ref{line}) with (\ref{gr-f-anz}) substituted
  is no more than the standard Schwarzschild line element
 in isotropic coordinates.

The   nonzero components of the Weitzenb\"{o}ck
connection (\ref{weiz2}) corresponding to the coframe
(\ref{gr-anz}) are
\begin{equation}\label{W-anz}
\oG_{0\hi}{}^0=\frac{\varphi'}{\varphi}\,\frac{x^\hm}\rho\d_{\hi\hm}\,,
\qquad
\oG_{\hi\hj}{}^\hk=\frac{\psi'}{\psi}\,\frac{x^\hm}\rho\,\d^\hk_\hi\d_{\hj\hm}\,.
\end{equation}
Consequently, the  independent nonzero components of the
Weitzenb\"{o}ck torsion tensor are
\begin{equation}\label{C-anz}
C_{0\hi}{}^0=\frac{\varphi'}{2\varphi }\,\frac
{x^\hm}\rho\d_{\hi\hm}\,,\qquad
 C_{\hi \hj}{}^\hk=\frac{\psi'}{2\psi }\,
 \frac {x^\hm}\rho(\d_{\hj\hm}\d^\hk_\hi-\d_{\hi\hm}\d^\hk_\hj)\,.
\end{equation}

The suitable ansatz for electromagnetic-type field
of a pointwise charge can be taken as
\begin{equation}\label{el-anz}
F_0{}^\hi=x^\hi f(\rho)\,, \qquad F_{0\hi}=\d_{\hi\hm}x^\hm
f(\rho)\psi^2\,.
\end{equation}

 The first Maxwell-type equation $F_{[ij;k]}=0$
 is satisfied  identically
when (\ref{C-anz}) and (\ref{el-anz})  are substituted. Indeed,
\begin{eqnarray}
F_{[0\hj,\hk]}&=&\frac 13(F_{0\hj,\hk}-F_{0\hk,\hj})\nonumber\\
&=&\d_{\hj\hm}(x^\hm f\psi^2)_{,\hk}-\d_{\hk\hm}(x^\hm
f\psi^2)_{,\hj} =0\,,
\end{eqnarray}
while
\begin{eqnarray}
&&C_{0\hj}{}^mF_{\hk m}+C_{\hj \hk}{}^mF_{0m}+C_{\hk 0}{}^mF_{\hj
m}\nonumber\\
&&\qquad =\frac{f\psi^2}2\left(\frac
{\varphi'}{\varphi}+\frac{\psi'}{\psi}\right)\frac{x^\hm
x^\hn}\rho(\d_{\hk\hm}\d_{\hj\hn}-\d_{\hk\hn}\d_{\hj\hm})=0\,.
\end{eqnarray}
The other components of the equation are zero because of
(anti)symmetry and independence of time.

As for the second Maxwell-type equation $F^i{}_{j;i}=0$, we
 observe that for $j\ne 0$ both sides vanish.
 For $j=0$, we substitute  (\ref{C-anz}) and (\ref{el-anz})
  into  (\ref{PreMax2-x}) to obtain
\begin{eqnarray}\label{an-max2}
f'\rho+3f=f\rho \left(\frac { \varphi'} \varphi-\frac{\psi'}\psi\right)\,.
\end{eqnarray}
This equation is straightforward integrated to
 \begin{equation}\label{mass-corr}
 f=\frac Q {\rho^3}\, \frac \varphi\psi=
 \frac Q {\rho^3}\, \frac{1-m/2\rho}{(1+m/2\rho)^3}\,,
 \end{equation}
 where $Q$ is  a constant of integration.
Consequently, the non-zero  field components are
 \begin{equation}\label{mass-corr1}
 F_0{}^\hi=x^\hi
 \frac Q {\rho^3}\, \frac{1-m/2\rho}{(1+m/2\rho)^3}\,.
 \end{equation}
 The Coulomb-type force acted on a test charge $q$ (of a  small mass)
takes the form
  \begin{equation}\label{mass-corr1x}
 F=
 \frac {Qq} {\rho^2}\, \frac{1-m/2\rho}{(1+m/2\rho)^3}\,.
 \end{equation}

The ordinary Cartesian radius $r$ is related to the isotropic radius $\rho$
 as
  \begin{equation}\label{iso-rad}
 r=\rho\left(1+\frac m{2\rho}\right)^2\,.
  \end{equation}
 Hence
 \begin{equation}\label{iso-rad1}
 \rho=\frac{r-m+\sqrt{r^2-2mr}}2\approx r\left(1-\frac mr\right)\,.
  \end{equation}
 Observe that the isotropic coordinates are defined only for $r>2m$.
 Consequently, the modernized Coulomb force takes the form
 \begin{equation}\label{mass-corr2}
 F=
 \frac {Qq} {r^2}\, \left(1-\frac {m^2}{4\rho^2}\right)\,.
 \end{equation}
 Finally,  in the Cartesian coordinates, the   mass-correction terms take the form
   \begin{eqnarray}\label{mass-corr3}
 F&=&Qq\,\frac{r-2m+\sqrt{r^2-2mr}}{4(r+\sqrt{r^2-2mr})^3}\nonumber\\&=&
 \frac {Qq} {r^2}\, \left(1-\frac {m^2}{4r^2}-\frac {m^3}{2r^3}+\cdots\right)
 \,.
 \end{eqnarray}
Observe the complicated form of the solution in the standard polar coordinates.
In fact,   the choice of  the isotropic coordinates can be viewed as a method
to solve the differential equation  $F^i{}_{j;i}=0$ exactly. Unfortunately,
this method is restricted to $r>2m$.

\begin{figure*}
\begin{tabular}{ccccc}
\begin{minipage}{2.55in}
\centering
\includegraphics[angle=270,width=2.55in]{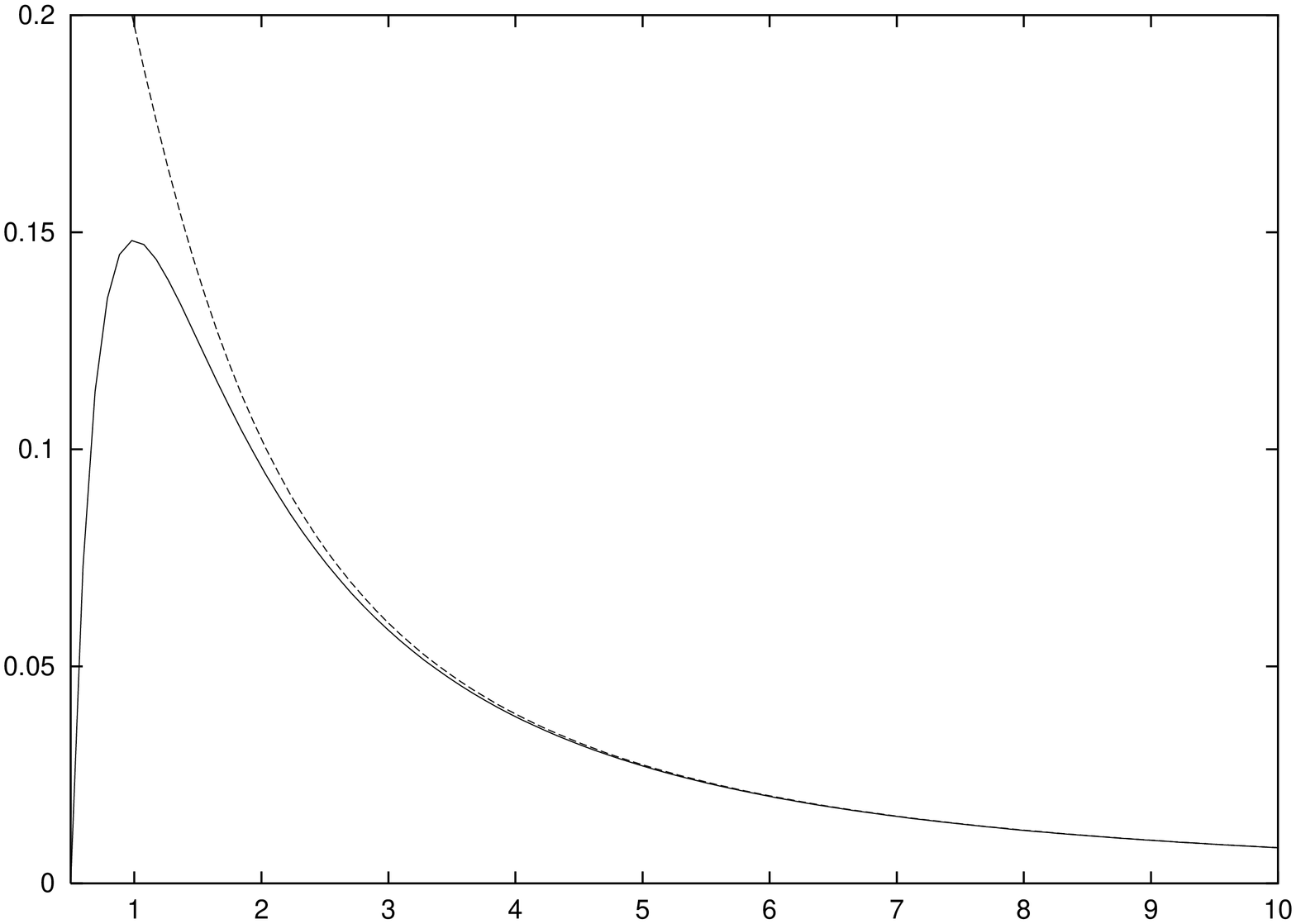}
\end{minipage}
&
\begin{minipage}{2.55in}
\centering
\includegraphics[angle=270,width=2.55in]{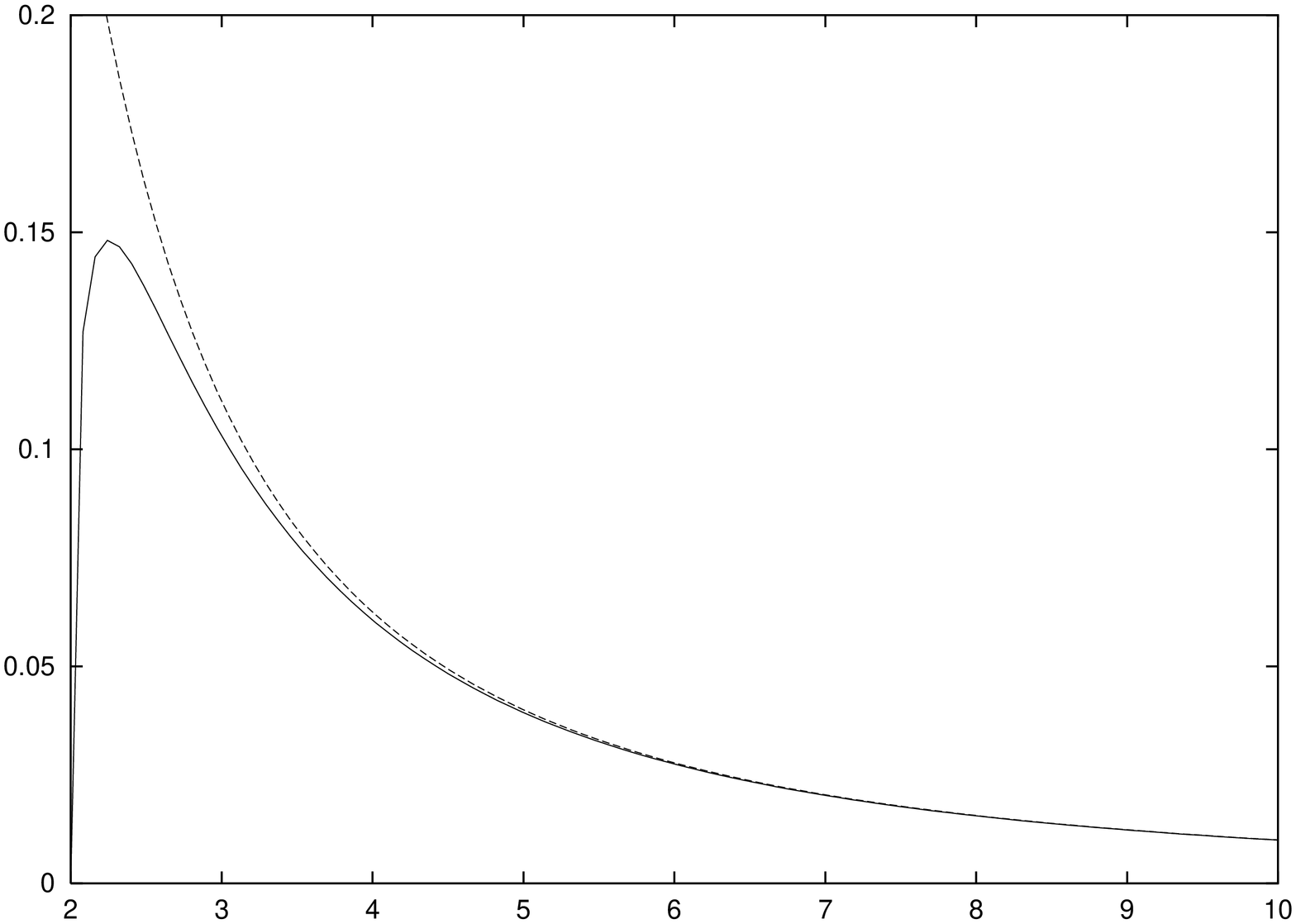}
\end{minipage}
\end{tabular}
\caption{The graphs represent the dependence of the force $F$ on $\rho/m$ and
$r/m$ correspondingly relative to the Coulomb force (the top lines). In both cases $F$ is given in the units of $Qq/m^2$.}
\label{fig:w_p2}
\end{figure*}
We depict the dependence of the force $F$ on the distances $\rho$ and $r$
on Fig. 2.
The graphs start from $\rho=m/2$ and $r=2m$ which are the minimal
possible value.  The deviation from the Coulomb values appears only for small
distances $\rho,r\sim m$. The maximal value of the force between two
 charged particles predicts as
\begin{equation}\label{force}
F\approx 0.15 \,{Qq}/{m^2}\,.
\end{equation}
 For two electrons, it gives
$F\approx 0.76\cdot 10^{86}N$.

\section{Outline of the model}

 \subsection{``Gauge geometry''}
 We construct a complete class of connections linear in the first order
 derivatives of the coframe field. It involves the standard
 Levi-Civita connection and the flat Weitzenb\"{o}ck
connection. For special choices of parameters, the torsion free
 and the metric compatible sub-families of connections emerge.
 Our main output is the identification of a sub-family of connections
 which are invariant under restricted local Lorentz transformations.
 The corresponding conditions are a set of eight first order equations.
 In the first order approximations the  rotations are described by  an
 antisymmetric matrix. In this case, the set of invariance conditions
 turns out to be the standard vacuum Maxwell system.

\subsection{Field equations}
We consider a Lagrangian of the matter-coframe
system
\begin{equation}
L=^{(cof)}L(\vt^\a,d\vt^\a)+^{(mat)}L(\psi,d\psi,\vt^\a)\,.
\end{equation}

In the standard consideration, the arbitrary variation of
the coframe field is assumed. The corresponding coframe
field equation is \cite{Itin:2001bp}
\begin{equation}\label{feq}
d*\F^\a=\T^\a\,,
\end{equation}
with the energy-momentum 3-form of the coframe-matter
system $\T=^{(cof)}\T+^{(mat)}\T$ in its right hand side.
Consequently (\ref{feq}) represents a system of 16
independent equations. It can be covariantly decomposed
to ten symmetric and six antisymmetric equations.
\begin{eqnarray}\label{feq1}
\vt^{(\b}\wedge d*\F^{\a)}&=&\vt^{(\b}\wedge \T^{\a)}\,\\
 \label{feq2}
\vt^{[\b}\wedge d*\F^{\a]}&=&\vt^{[\b}\wedge \T^{\a]}\,,
\end{eqnarray}
 Consider the pure coframe system, and restrict
 to the viable case $\rho_1=0$.
The explicit calculations, see \cite{Itin:1999zs}, show that both sides
 of (\ref{feq2}) involve the leading coefficient $2\rho_2+\rho_3$.
 Consequently, for $2\rho_2+\rho_3=0$, only ten independent
 field equations (\ref{feq1}) remain. Certainly, this symmetric
 system is equivalent to Einstein equation. In this case,
 the coframe field is  defined only
 up to arbitrary Lorentz transformations.
 For the alternative models with $2\rho_2+\rho_3\ne 0$,
 (\ref{feq1}-\ref{feq2}) is a well posed system of 16
 hyperbolic PDE for 16 independent variables. The coframe field
 is defined uniquely (up to global transformations).

In the approach proposed in the current paper, only the symmetric
 variations of the coframe field are independent.
 The antisymmetric variations are constrained by the system
 (\ref{Max-Rie}). Since the variation derivative of the Lagrangian
 is covariantly decomposed to the symmetric and antisymmetric
 parts,  only one field equation,
 namely the symmetric one, (\ref{feq1}), is derived from
 the symmetric variation of the action.  The second field equation
 comes from the invariance of the Lagrangian. Consequently
 our system of equations
 includes  16 independent equations for 16 independent variables
 \begin{eqnarray}\label{feq3}
&&\vt^{(\b}\wedge d*\F^{\a)}=\vt^{(\b}\wedge \T^{\a)}\,\\
 \label{feq4}
&&F_{[ij;k]}=0\,, \qquad F^i{}_{j;i}=0\,.
\end{eqnarray}
 The first (``gravitational'') equation has to give a class of 
 orthonormal coframes, i.e., a solution that valid 
 up to arbitrary transformations of the coframe. 
 . The second
 (``electromagnetic'') equation has to define uniquely
 (up to global transformations) the field of local rotations.
 Although, the well-posedness of the system (\ref{feq3}-\ref{feq4})
 requires a special consideration, it is rather reasonable. Indeed:

 (i) In the first order approximation \cite{Itin:2004ig}, the coframe field
 is decomposed to a sum of symmetric and antisymmetric fields.
 For $\rho_1=0$,  also the system of field equations is separated ---
 (\ref{feq3}) involves only the symmetric part, while (\ref{feq2})
 describes the dynamics of the antisymmetric part.
 Moreover, in this approximation, (\ref{feq2}) takes the form
 $\square F_{ij}=0$, which is a consequence of  (\ref{feq4}).

 (ii) For a spherical-symmetric ansatz, (\ref{feq3}) is equivalent to the
 vacuum Einstein equation, i.e., determines uniquely a class of
 orthonormal coframes. When the solution is substituted in  (\ref{feq4}) also
 the field $F_{ij}$ is determined uniquely (see the previous section).

 (iii) The equation (\ref{feq4}) involves the torsion tensor $C_{ijk}$
 only linearly.
 Thus, it can be viewed as a linear
 combination of the derivatives $F_{ij,k}$. When these combinations
 are substituted into (\ref{feq3}), the corresponding equation is a
 system of second order quasi-linear hyperbolic PDE for $F_{ij}$.
 \subsection{Pure ``gravity'' sector}
 
 Let the local Lorentz rotations $X_{\a\b}$ (and, consequently, 
 their  approximations $F_{ij}$) are assumed to be independent on a point.
 In this case, the field equations (\ref{feq4}) are satisfied.
 The remain 10
 equations (\ref{feq3}) is an under-defined system for 16 independent
 components of the coframe field. The metric tensor, however, is
 defined uniquely. In particular, with a mentioned restriction
 $\rho_1=0$, the system has a unique spherical-symmetric solution
 which corresponds to the Schwarzschild metric.

 Moreover, the coframe field model for gravity has even some advantage
 relative to the standard metric GR. It is well known that an
 energy-momentum tensor cannot be constructed covariantly from
 the metric tensor and its first order derivatives. Consequently,
 the energy of a gravity field cannot be defined in GR.
 Alternatively, for a coframe field, the corresponding tensor is defined
 \cite{Itin:2001bp}. This tensor is not local Lorentz in Varian and so cannot
 be prolongated into the standard metric GR. However, in the model proposed
 here, the arbitrary local Lorentz transformations are not acceptable.
 Moreover, these transformations are
 governed by their own field equations.
 \subsection{Pure ``electromagnetic''  sector}
 Let the coframe field be defined up to a local pseudo-rotation. Thus the
 metric is defined uniquely.
 Moreover, let a representative coframe be chosen.
 The set of invariance conditions now is a system of eight first order
 equations for six independent variables $L_{\a\b}$. In fact,
 we have here a linear field model for $L_{\a\b}$, which turns to a
 non-linear field model for $X_{\a\b}$, where
 \begin{equation}
 L_{\a\b}=\eta_{\a\b}+X_{\a\b}+\mathcal{O}(X^2)\,.
 \end{equation}
 A similar non-linear extension of
 electrodynamics based on orthonormal tensor
 was recently discussed  \cite{Coll:2003ji}.
 It can give an alternative to the
 Born-Infeld electrodynamics, which turns  recently to a popular model in
 strings theory.

 Recall, that in the current paper we consider only the first order
 approximation of the pseudo-orthonormal tensor $L_{\a\b}$.
 The corresponding invariance conditions are
 \begin{eqnarray}\label{PreMax-x1}
K_{[ijk]}=0\,, \qquad K^m{}_{im}=0\,,
\end{eqnarray}
or, after a redefinition of the variables,
  \begin{eqnarray}\label{PreMax-xx}
F_{[ij;k]}=0\,, \qquad F^m{}_{i;m}=0\,,
\end{eqnarray}
 where the covariant derivative is taken relative to the flat Weitzenb\"{o}ck
connection.
For a holonomic coframe field, these equations take the form of the
 standard Maxwell system
   \begin{eqnarray}\label{PreMax-xxx}
F_{[ij,k]}=0\,, \qquad F^m{}_{i,m}=0\,.
\end{eqnarray}
Observe the main properties of the derived field equations:
 \begin{itemize}
 \item[(i)] The (approximated) Maxwell-type system is not introduced by hand, but
 emerges in a natural way as a set of invariance condition for a geometry
 and for a viable Lagrangian.
  \item[(ii)] The complete field equation is nonlinear, what is
 in a correspondence with the string model consequences.
\item[(iii)] The spherical symmetric solution is bounded near
 the Schwarzschild radius. Further off, it is close to the
 Coulomb field.
\item[(iv)] It is well known that in the models with an asymmetric
 connection, the torsion contribution is an obstacle for the
 standard definition of the potentials,
 $F_{ij}=\partial_i A_j-\partial_j A_i$.
 For torsion of a restricted type,  the  potentials can be defined, 
 however, as
 \begin{equation}
 F_{ij}=\partial_i A_j-\partial_j A_i+2T^k{}_{ij}A_k
 \end{equation}
 and the modified gauge transformations
are acceptable \cite{Hojman:1978yz}.
 \item[(v)] In the standard Maxwell theory a charge is not connected
 to a mass. Thus, unphysical massless charges are not forbidden.
 In our construction, the field of pseudo-rotations can be degenerate
 only in the singularities of the coframe field (metric).
 With this requirement, the charges are necessary massive.
 Certainly, the field of rotations can be regular in the
 whole space even the coframes are singular. In this case,
 we have uncharged masses.
 \item[(vi)] A nonzero constant field $F_{ij}$  is another
 unphysical solution
 of the Maxwell system, which is usually removed by the boundary
 conditions.
 The modified field equation (\ref{PreMax-xx}) does not have
 such a solution.
  \item[(vii)] In the first order approximation, the
  coframe can be covariantly decomposed in a sum of its
  symmetric and antisymmetric parts.  In this case, the
  difference between coordinate and coframe indices can be
  neglected. Consequently, we can write
  \begin{equation}\label{fin1}
  \vt_{ij}=\vt_{(ij)}+\vt_{[ij]}=g_{ij}+\frac 1\beta F_{ij}\,.
  \end{equation}
The free parameter $\beta$ has a dimension of a field
strength. Note that the structure of (\ref{fin1}) is
similar to a combination appearing in the Born-Infeld
electrodynamics, where $\beta$ is a maximal
electromagnetic field. From our spherical symmetric
solution $\beta\approx m^2/(0.15 {q})$. At this stage, it
is only a speculation because $\beta$ has sense only in
higher order approximations.
 \item[(viii)] In this paper, we have considered only the
 vacuum case. A complete model has to include a Lagrangian
 for fermions. Due to Fock and Dirac, the fermionic
 Lagrangian on a curved space is constructed with the spin
 connections. A remarkable fact that this Dirac connection
 is proportional to the flat connection of Weitzenb\"{o}ck. 
 Consequently, the covariant derivative is the same as in 
 (\ref{PreMax-xx}).
 
\item[(ix)] Qualitatively, the coframe structure in our model
can be described as a "paramagnetic substance". The
gravitational sector establishes the coframe at every
point, i.e., the lengths of four covectors and the angles
between them. The coframes do not interact with one another
and are randomly oriented. The metric on the manifold is
defined uniquely. The electromagnetic-type sector orders the
coframes attached at different points in a smooth coframe
field. Different type of ordering correspond to different
connections from the six parametric family described above.
 \end{itemize}

\section{Results}
On a manifold endowed with a coframe (vierbein, tetrad)
field, we constructed a most general linear asymmetric
connection. Besides the torsion-free and metric-compatible
connections, we identified a subset of Maxwell-compatible
connections. The vacuum Maxwell-type equations emerges as the
conditions for invariance of the connection under Lorentz
transformations of the coframe.

We derive the same equations as invariant conditions for a
viable coframe Lagrangian which has the Schwarzschild
solution even being alternative to GR.

In our approach, the curved space Maxwell-type equations emerges
formally the same as in the standard Maxwell-Einstein
system.   The covariant derivatives is different, however,
in proposed model they are taken relative to the flat connection.
 The result is a different behavior of the
electromagnetic field near the Schwarzschild radius.
Moreover, the model predicts an upper bound for the
electromagnetic-type field of a source with given charge and
mass. Note that although the gravitational field of a
charged source remains independent on the charge, it is
only a result of the approximation used. Already in the
second approximation, the metric tensor is not invariant
under  the first order approximate Lorentz transformations
and a quadratic correction, $F_i{}^mF_{mj}$, emerges in the
metric tensor.

 In the current paper, we have treated only the vacuum case. 
 The problem of the 
 sources for the proposed Maxwell-type field requires additional 
 investigations. In particular, they can appear from a viable modification 
 of Dirac field on a coframe background. 
\section*{Acknowledgment}
 A part of this paper was written during my  visit to MIT.
 I am grateful to  Roman Jackiw for warm hospitality and  fruitful
 discussions.
 I thank Shmuel Kaniel,  Friedrich Hehl,  Jacob  Bekenstein and
 David Kazhdan for valuable remarks. Many thanks are due to the referee 
 for most constructive suggestions. 

\end{document}